\documentclass[11pt,twocolumn]{article}
\usepackage[margin=1in]{geometry}
\usepackage{amsmath,amsthm,amssymb,amsfonts,braket}

\usepackage{graphicx}
\usepackage{subcaption}

\usepackage{datetime}
\usepackage{balance}
\usepackage{color}
\usepackage{bm,upgreek}

\usepackage[normalem]{ulem}
\usepackage{marginnote}

\newcommand{\stkout}[1]{\ifmmode\text{\sout{\ensuremath{#1}}}\else\sout{#1}\fi}

\begin{document}

\title{Diffusive Persistence on Disordered Lattices and Random Networks}
\author{Omar Malik$^{*1,2}$, Melinda Varga$^{3}$, Alaa Moussawi$^{1,2}$, David Hunt$^{1,2}$, Boleslaw Szymanski$^{1,2,4}$,\\ Zoltan Toroczkai$^{3}$, Gyorgy Korniss$^{*1,2}$}

\date{
{\small $^1$ Department of Physics, Applied Physics, and Astronomy, Rensselaer Polytechnic Institute, Troy, New York 12180,
USA.\\%
$^2$ Network Science and Technology Center, Rensselaer Polytechnic Institute, Troy, New York 12180, USA.\\
$^3$ Department of Physics and Astronomy, University of Notre Dame, Notre Dame, Indiana
46556, USA.\\%
$^4$ Department of Computer Science, Rensselaer Polytechnic Institute, Troy, New York 12180,
USA.\\
$^*$ Corresponding authors: drundsmas@gmail.com, korniss@rpi.edu}\\[2ex]%
}


\maketitle

\begin{abstract}
To better understand the temporal characteristics and the lifetime of fluctuations in stochastic processes in networks, we investigated diffusive persistence in various graphs. Global diffusive persistence is defined as the fraction of nodes for which the diffusive field at a site (or node) has not changed sign up to time $t$ (or in general, that the node remained active/inactive in discrete models).
Here we investigate disordered and random networks
and show that the behavior of the persistence depends on the topology of the network. In two-dimensional (2D) disordered networks, we find that above the percolation threshold diffusive persistence scales similarly as in the original 2D regular lattice, according to a power law $P(t,L)\sim t^{-\theta}$ with an exponent $\theta \simeq 0.186$, in the limit of large linear system size $L$. At the percolation threshold, however, the scaling exponent changes to $\theta \simeq 0.141$, as the result of the interplay of diffusive persistence and the underlying structural transition in the disordered lattice at the percolation threshold.
Moreover, studying finite-size effects for 2D lattices at and above the percolation threshold, we find that at the percolation threshold, the long-time asymptotic value obeys a power-law $P(t,L)\sim L^{-z\theta}$ with $z\simeq 2.86$ instead of the value of $z=2$ normally associated with finite-size effects on 2D regular lattices.
In contrast, we observe that in random networks without a local regular structure, such as Erd\H{o}s-R\'enyi networks, no simple power-law scaling behavior exists above the percolation threshold.
\end{abstract}

\section*{Introduction}

Persistence type questions \cite{Bray2013} in general refer to a certain event not happening, or a certain property or a pattern surviving for a given period of time. Examples include: the probability that a noise spike  created at $t=0$ in a telegraph line diminishes after time $t$; the probability of an earthquake of a given size or larger not happening for $n$ consecutive years; the probability  of survival of domains in magnetic systems up to time $t$; how long a pandemic will last, etc.~\cite{SALCEDOSANZ20221}. In other words, the study of persistence refers to the description of the duration of the excursions taken by some observable in a stochastic process from some chosen threshold  \cite{Fuchs_2008}, often taken as the mean for this process. The observable being studied may be global, such as a bulk order parameter, or local, such as the value of a field at a node. Despite the apparent simplicity and the practical importance of these questions, only a handful of exact results are known.

Probably the earliest accounts on persistence-like questions go as far back as 1887 when Bertrand \cite{Feller} proved his ballot theorem. In the ballot problem, two candidates, $B$ and $G$ score a total of $p+q$ votes with $B$ scoring $p$ votes and $G$ scoring $q$ votes. Given we know the end result, i.e., for example $B$  won ($p > q$), what is the probability that during the counting of votes $B$ always was in the lead? The answer is $(p-q)/(p+q)$.
This is easily solvable because it is a one-body (single point) Markovian process.   Another one-body problem is that of the randomly accelerated particle described by the Langevin equation $d^2 \chi / dt^2 = \eta(t)$, where $\chi(t)$ is the position of the particle at time $t$ and $\eta(t)$ is a solely time-dependent, white noise. Here we are asking: What is the probability $P(t)$ that the particle does not cross $\chi = 0$ up to time $t$? Being no longer Markovian (two-step memory due to the second-order derivative), it is harder to solve, with the answer being $P(t) \sim t^{-1/4}$ \cite{Sinai1992, Burkhardt1993}.

Many-body persistence problems, i.e., processes where there are two or more coupled variables evolving in time, are typically non-tractable analytically, save for a few cases. A large family of such problems studies the persistence properties of stochastic processes on regular lattices. Examples include diffusion \cite{satya96,newman98}, interface fluctuations \cite{Krug_PRE1997,Toro_PRE1999}, magnetic and reaction-diffusion systems \cite{Derrida_JPA2004,Derrida_PRL1995,Howard_JPA1998}, and contact processes \cite{Fuchs_2008,Grassberger_JSM2009}. For an Ising or Voter model \cite{Derrida_JPA2004,Derrida_PRL1995,Howard_JPA1998}, of interest is the probability that the local state variable has never switched by time $t$.
The persistence probability is also of particular interest in non-equilibrium or disordered systems such as spin-relaxation in the Ising model \cite{derrida96,Newman1999} and the Blume-Capel model \cite{Silva2004}, as well as the persistence of the bulk order parameter in inhomogeneous magnetic systems with defects \cite{Pleimling2005}.
In experiments, persistence probabilities were measured in a variety
of systems including in breath figures \cite{MBBGY}, in twisted nematic
liquid crystals \cite{YPMS}, soap bubbles \cite{TZSS} and in dense
spin-polarized noble gases (He3 and X-129) \cite{TPNO}.

One of the most studied many-body persistence problems is diffusive persistence. In the classic version and using continuous formulation, we monitor the relaxation of a field $\Phi(\bm{x},t)$, $\bm{x} \in {\cal D} \subseteq \mathbb{R}^{d}$ through the diffusion equation

\begin{align}
    \partial_{t}\phi = D \nabla^{2} \phi,
\end{align}
with initial condition $\Phi({\bf x},0) = \psi({\bf x})$, where $\psi({\bf x})$ is an uncorrelated random field taken from a Gaussian
distribution $G[\psi] \sim e^{-\frac{1}{2\Delta} \int d^d x \; \psi({\bf x})^2}$ with zero mean and variance $\Delta$. Let $P_{\bm{x}_{0}}(t)$ denote the {\em diffusive persistence probability} (DPP), i.e, the probability that the field at some $\bm{x}_{0}$ did not change sign up to time $t$, i.e.
\begin{equation}
P_{\bm{x}_{0}}(t) = \mbox{Prob}\{\psi(\bm{x}_{0})\Phi(\bm{x}_{0},\tau) > 0,\;\forall\;\tau \in [0,t)\}.
\end{equation}
The goal is to compute $P_{\bm{x}_{0}}(t)$ and its asymptotic behavior at large times. Both numerically and analytically, the problem is studied in the discrete formulation on the lattice $\mathbb{Z}^{d}$, using a discretized version of the diffusion equation. The numerical observation is that for the $d$-dimensional infinite square lattice, the asymptotic behavior is a power-law
\begin{equation}
P_{\bm{x}_{0}}(t) \sim t^{-\theta}\;, \label{pow}
\end{equation}
where $\theta$, called the {\em persistence exponent} depends on the dimension of the lattice
$\theta = \theta_{d}$. Clearly, this problem is translationally invariant and thus the DPP does not depend on $\bm{x}_{0}$. In particular, $\theta_{1} = 0.1207\pm 0.0005$ and $\theta_{2} = 0.1875\pm 0.0010$ \cite{satya96,newman98}.

The asymptotic power-law behavior Eq.~\eqref{pow} appears to hold for a wide variety of other systems \cite{hari98}, beyond the diffusion equation, with exponents $\theta$ that are typically not {\em simple rational} numbers.
Despite the simple conceptual definition of the persistence probability (related to the first-passage time distribution \cite{Redner} for zero crossings of the local field variables) there are typically no exactly known values for $\theta$ for many-body persistence problems, apart from some exceptional cases. One such exceptional case is the DPP exponent in 2D above, which is a simple rational number, with $\theta_{2} = 3/16$~\cite{schehr2018}  derived only recently by Poplavskyi and Schehr exploiting a connection to Kac random polynomials and the truncated orthogonal ensemble of random matrices. This is a peculiar case as typically exact results are obtained for lower dimensions more easily. However, there are still no exact results for $\theta_{1}$ and the method in \cite{schehr2018} does not seem to be easily modifiable for the 1D case.

Studies of persistence have mostly focused on stochastic dynamics taking place over homogeneous, translationally invariant systems embedded in a metric space and in the infinite system size limit. This is a rather idealized setup, with limited every-day practical applicability. However, answering persistence type questions for real-world systems, and in particular, networks, would have significant importance. For example,  in critical infrastructure networks, persistence can be defined as the probability that a local region remains operational up to time $t$. In influencing and opinion dynamics in social networks,  persistence can be defined as the probability that certain nodes or network regions have not changed opinion since the beginning of a ``campaign", etc. Such applications raise several interesting questions: Given a stochastic dynamics over a network, and a network observable, what are the network properties that most influence the persistence properties of that observable? What are the fundamental differences in persistence properties taking place over networks embedded in a metric space (spatial networks) instead of on general graphs?

In order to attempt answering these questions, in this paper we present a set of computational studies of diffusive persistence on various regular and disordered network structures, including networks obtained from bond percolation, 2D random geometric graphs, $k$-regular random graphs, the $k$-th power of a circle and Erd\H{o}s-R\'{e}nyi (ER) random graphs. The paper concludes with a discussion of these numerical observations.

\section*{Diffusive persistence on networks}

 We define a diffusive field variable $\psi_i$ for the $i^{\text{th}}$ node of the network. The value of this variable for each node in the network is initialized by sampling from a normal distribution with 0 mean and a standard deviation of 1. Following previous work \cite{satya96, newman98}, the discretized diffusion equation that we will study is
\begin{align}
\psi_i(t + \Delta t) - \psi_i(t) =  - \alpha\Delta t \sum_j A_{ij}(\psi_i(t) - \psi_j(t)),
\end{align}
where $A_{ij}$ is the adjacency matrix ($A_{ij} \in \{0,1\} $) associated with the network and the RHS of the equation represents the graph Laplacian operator. For 2D lattices we choose $\alpha = 1$ and $\Delta t = \frac{1}{8}$.

Since, in general, we no longer have translational invariance, we have to distinguish local and global persistence
measures. The local persistence $P_{i}(t)$ at site $i$ is defined in the same way as before, and the global persistence probability, $P(t, L)$, is defined as the fraction of nodes for which $\psi_i(t)$ has not changed sign at time $t$ on a lattice with $N = L^2$ nodes. Clearly, $P_{i}(t)$ can depend on the site $i$, in general, and we are not studying this quantity here, we focus exclusively on the global persistence probability $P(t, L)$.

For every network configuration we generate 100 samples and simulate the diffusion equation with the value of the field at each node sampled from a normal distribution with zero mean and unit variance.

\subsection*{Bond-percolating disordered lattices}

The percolation process on a lattice may be understood as removing either the connections between sites (bond-percolation) or the sites themselves (site percolation) \cite{stauffer92}. The process is characterized by $\phi$, where $1-\phi$ is the probability of removal. Our 2D regular lattices in this work (before edge removal) are 2D square lattices with four nearest neighbors.
Percolation on a 2D lattice creates a fractal structure, with the percolation threshold, $\phi_c$, marking the point at which an infinite cluster first appears when performing the reverse process (adding the bonds randomly) \cite{isichenko92}. The fractal dimension of the lattice for $\phi\ge\phi_c$ is 2, while below the percolation threshold it is 1.896. Thus the topology of the lattice undergoes a phase transition as it becomes disordered.

The effect of this change in topology naturally affects any process on the lattice. Diffusion on lattices, for example, is associated with power-law behavior. When the 2D lattice is above the percolation threshold, the average distance traversed in a random walk behaves as $R \propto t^{\frac{1}{2}}$. However, when the network is at its {\em site} percolation threshold of $\phi_c^{site} \simeq 0.59$, the behavior changes to $R \propto t^{\frac{1}{D'}}$, where $D' = 2.85 \pm 0.05$ \cite{havlin1983}.

To create disordered networks for our persistence studies, we iterate through the edge list and randomly remove edges with probability $1-\phi$. We then repeat the diffusive process on the giant component of the resulting disordered network. The bond-percolation threshold of 2D square lattices is $\phi_{c} = 0.5$, and we vary $\phi$ in the range $\left[0.5, 1\right]$.

\begin{figure}
	\centering
	\includegraphics[width=\linewidth]{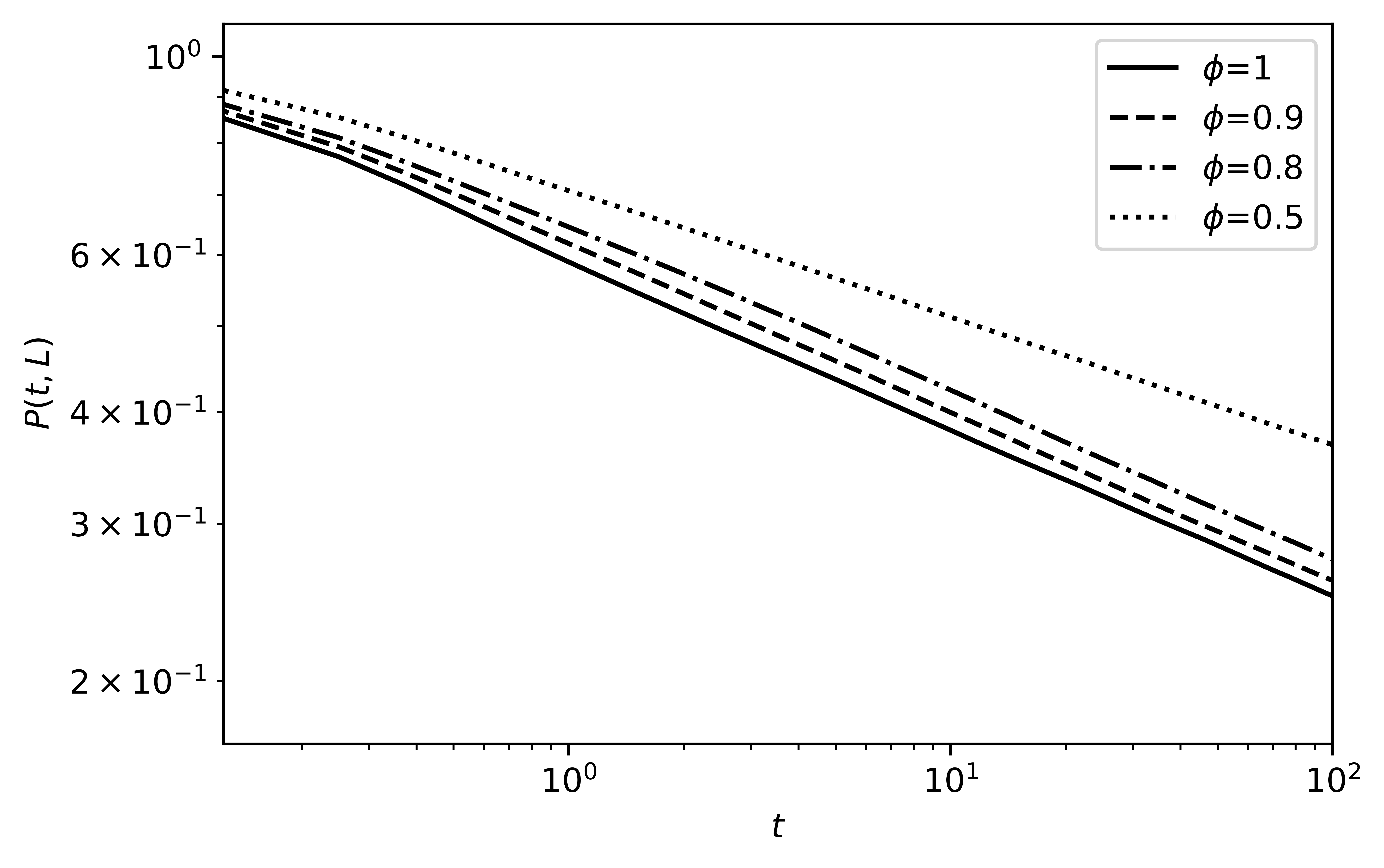}
	\label{fig:2d}
	\caption{Persistence on a 2D regular and disordered lattices with $L=100$, where $1-\phi$ is the edge-removal probability. The diffusive persistence exhibits a power-law behavior. However, as edges are removed from the network, the power-law exponent changes. At the percolation threshold ($\phi=0.5$), the power-law exponent is $\theta = 0.141~\pm~5.3\times 10^{-5}$.}\label{fig:2dper}
\end{figure}

\begin{table}[]
\centering
\begin{tabular}{|l|l|}
\hline
$\phi$ & $\theta$ \\ \hline
1 & $0.186 \pm 1.4 \times 10^{-4}$  \\ \hline
0.9 & $0.189 \pm 1.6 \times 10^{-4}$ \\ \hline
0.8 & $0.189 \pm 1.4 \times 10^{-4}$ \\ \hline
0.5 &  $0.141 \pm 5.3 \times 10^{-5}$\\ \hline
\end{tabular}
\caption{Values of the power-law exponent, $\theta$, associated with diffusive persistence on 2D lattices with $L=100$ nodes as a function of $\phi$, where $1-\phi$ is the edge-removal probability.}
\label{tab:2d_per}
\end{table}

Fig.~\ref{fig:2dper} shows the persistence probability $P(t, L)$ as diffusion proceeds on a 2D network with a $10^4$ nodes. For the fully-intact network (i.e $\phi = 1$) we recover the known power-law behavior Eq.~\eqref{pow} with $\theta = 0.186 \pm 1.4 \times 10^{-4}$ (the exact result is $3/16 = 0.1875$). For different values of $\phi$ we still observe power-law behavior but with a different persistence exponent at the percolation threshold. In order to characterize the change in $\theta$ as a function of $\phi$ we fit the different persistence probability decay curves. The results are shown in Tab.~\ref{tab:2d_per}. The exponent undergoes a dramatic shift at the percolation threshold. The persistence at the percolation threshold follows a power-law with novel exponent $\theta = 0.141~\pm~5.3 \times 10^{-5}$.

Away from the percolation threshold the power-law exponent maintains its value of approximately $\theta\simeq 0.189$ for $\phi > 0.8$. The region $0.5 < \phi < 0.8$ acts as a slow crossover region, where the decay is
not a clean power-law, so we could not associate $\theta$ values in this regime.

\subsubsection*{Finite-size behavior }
\begin{figure*}
	\centering
	\begin{subfigure}[l]{0.45\linewidth}
		\includegraphics[width=\textwidth]{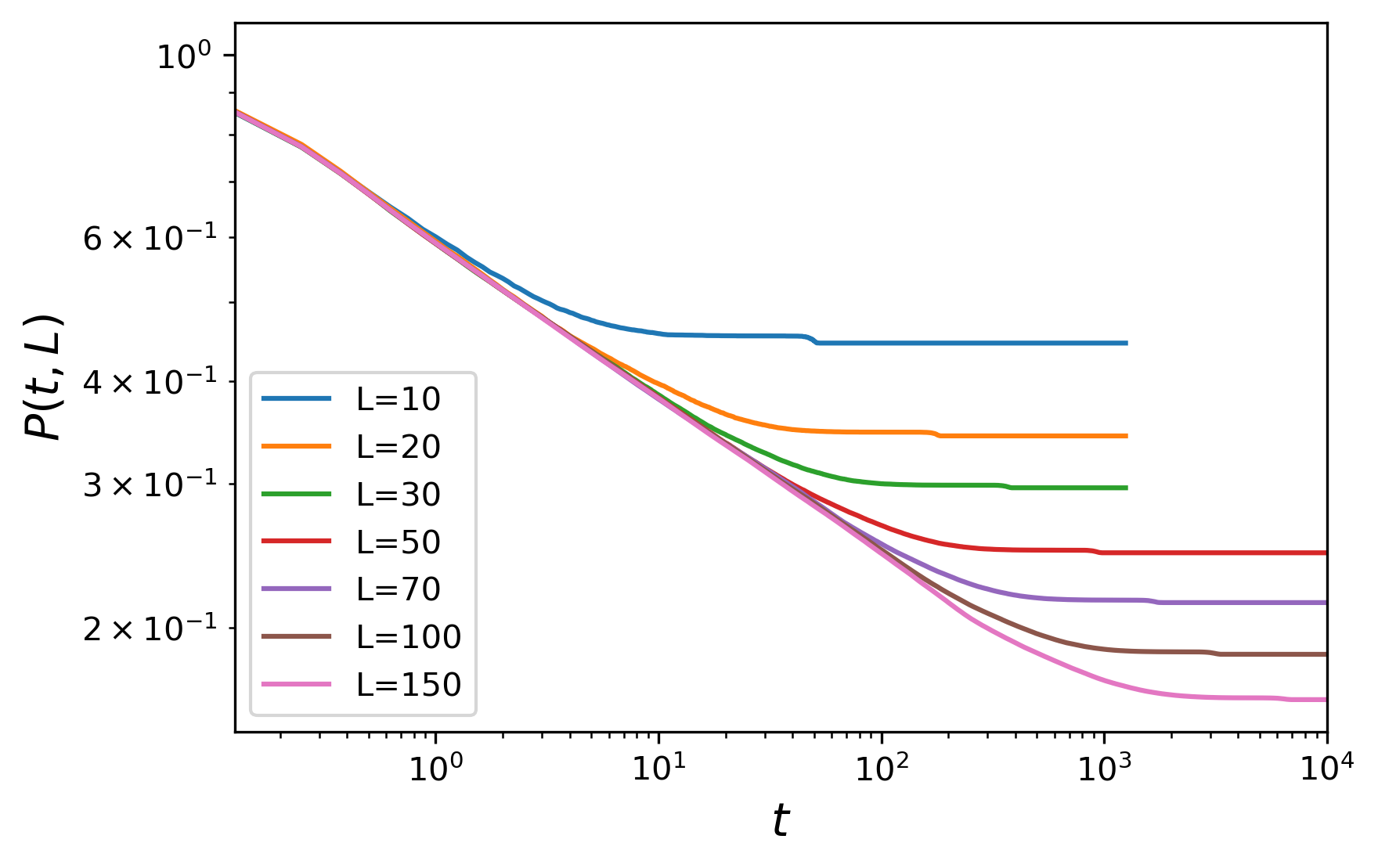}
		\caption{}
		\label{fig:2d_fin_size_1}
	\end{subfigure}
	\begin{subfigure}[l]{0.45\linewidth}
		\includegraphics[width=\textwidth]{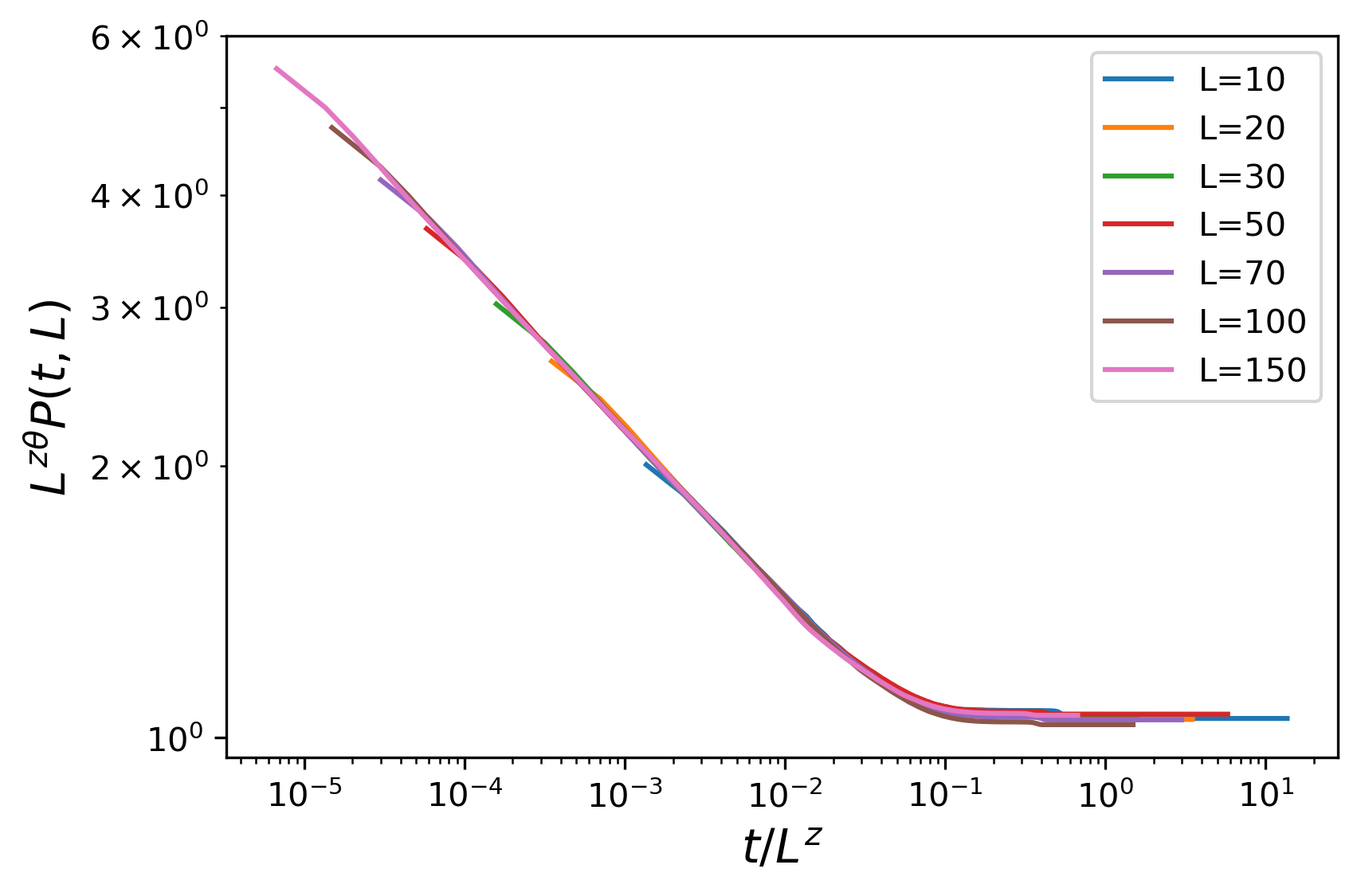}
		\caption{}
		\label{fig:2d_scaling_phi_1}
	\end{subfigure}
	~ 
	\caption{(a) Finite-size effects on 2D regular lattices with $\phi=1$. (b) Scaling behavior of 2D regular lattices for $\phi = 1$ with $\theta = 0.186$ and $z = 1.99$ (reproducing earlier results of Ref.~\cite{manoj2000}).}
\end{figure*}

\begin{figure*}
	\centering
	\begin{subfigure}[l]{0.45\linewidth}
		\includegraphics[width=\textwidth]{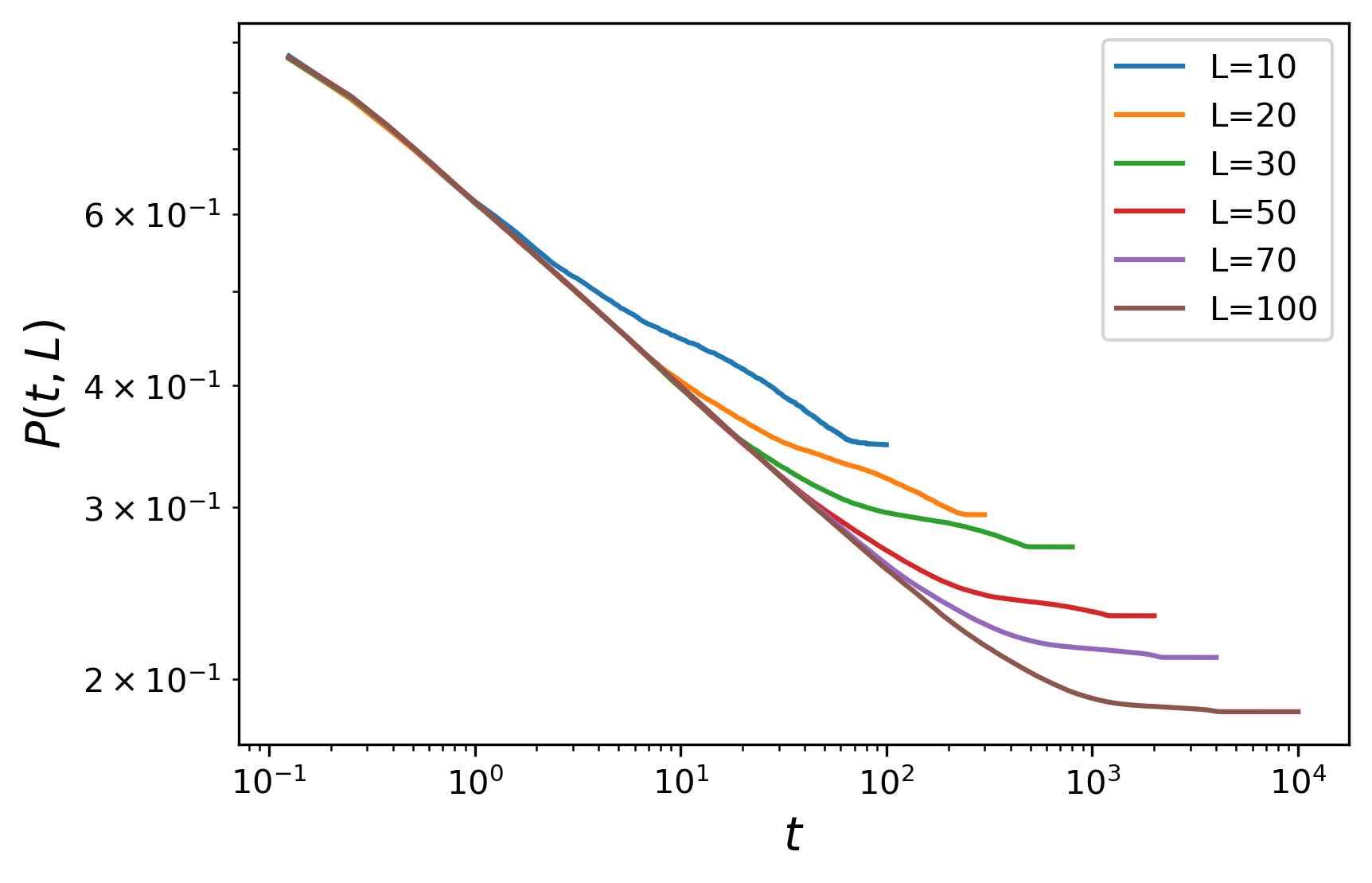}
		\caption{}
		\label{fig:2d_fin_size_0_9}
	\end{subfigure}
	\begin{subfigure}[l]{0.45\linewidth}
		\includegraphics[width=\textwidth]{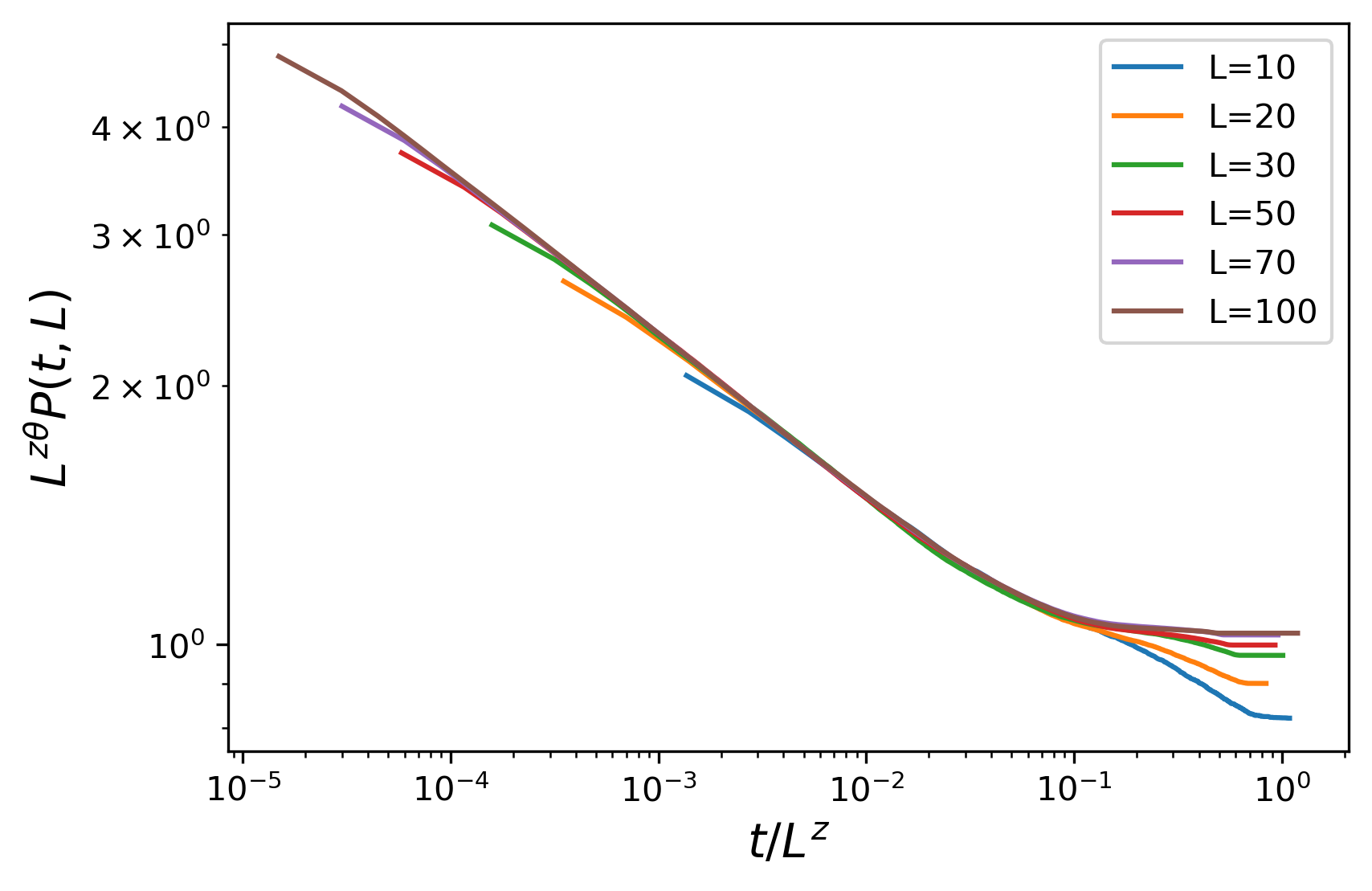}
		\caption{}
		\label{fig:2d_scaling_phi_0_9}
	\end{subfigure}
	~ 
	\caption{(a) Finite-size effects on 2D disordered lattices with $\phi=0.9$. (b) Scaling behavior of 2D disordered lattices for $\phi = 0.9$ with $\theta = 0.189$ and $z = 1.99$.}
\end{figure*}

\begin{figure*}
	\centering
	\begin{subfigure}[l]{0.45\linewidth}
		\includegraphics[width=\textwidth]{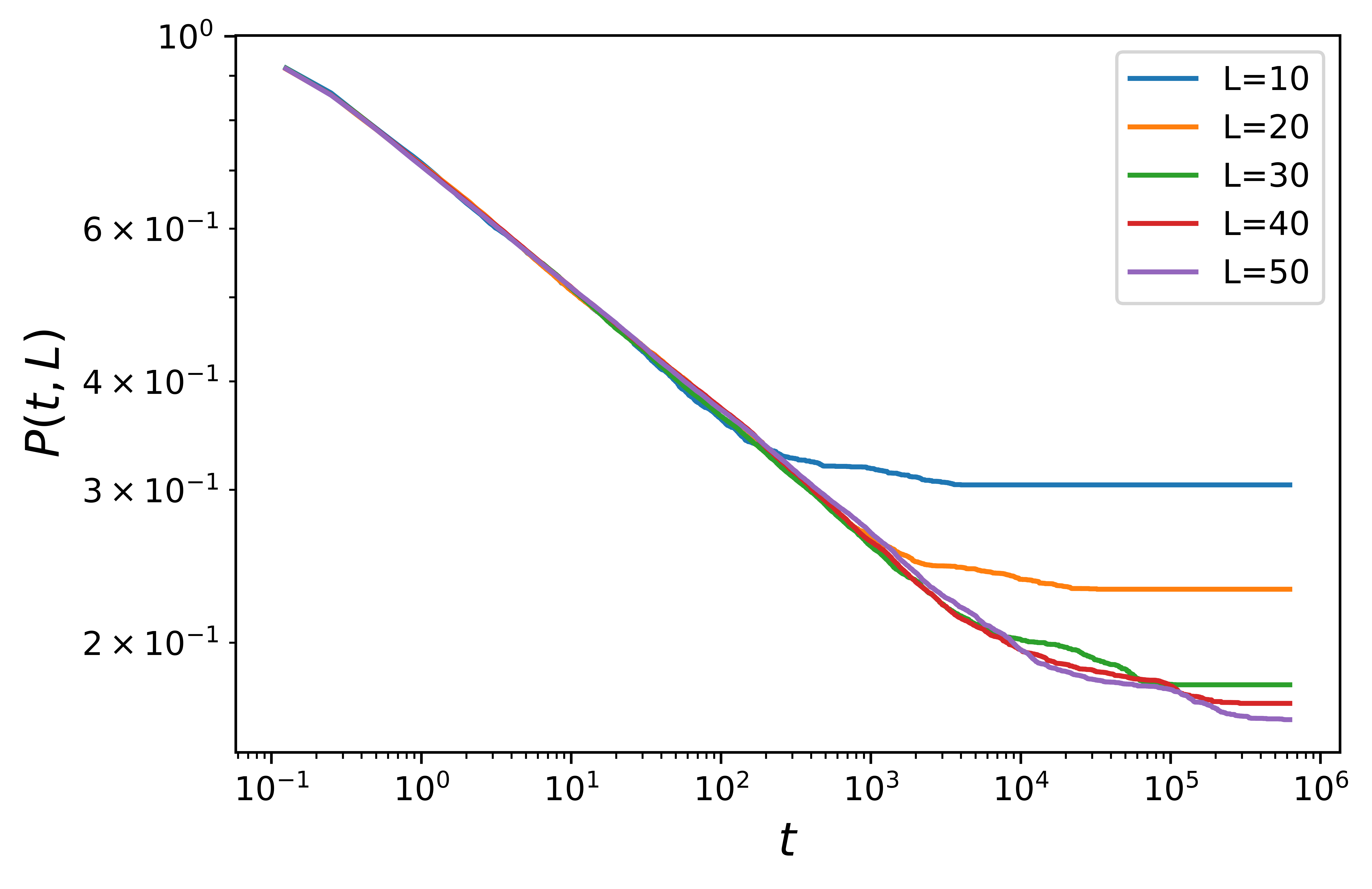}
		\caption{}
		\label{fig:2d_fin_size_phi_0_5}
	\end{subfigure}
	\begin{subfigure}[l]{0.45\linewidth}
		\includegraphics[width=\textwidth]{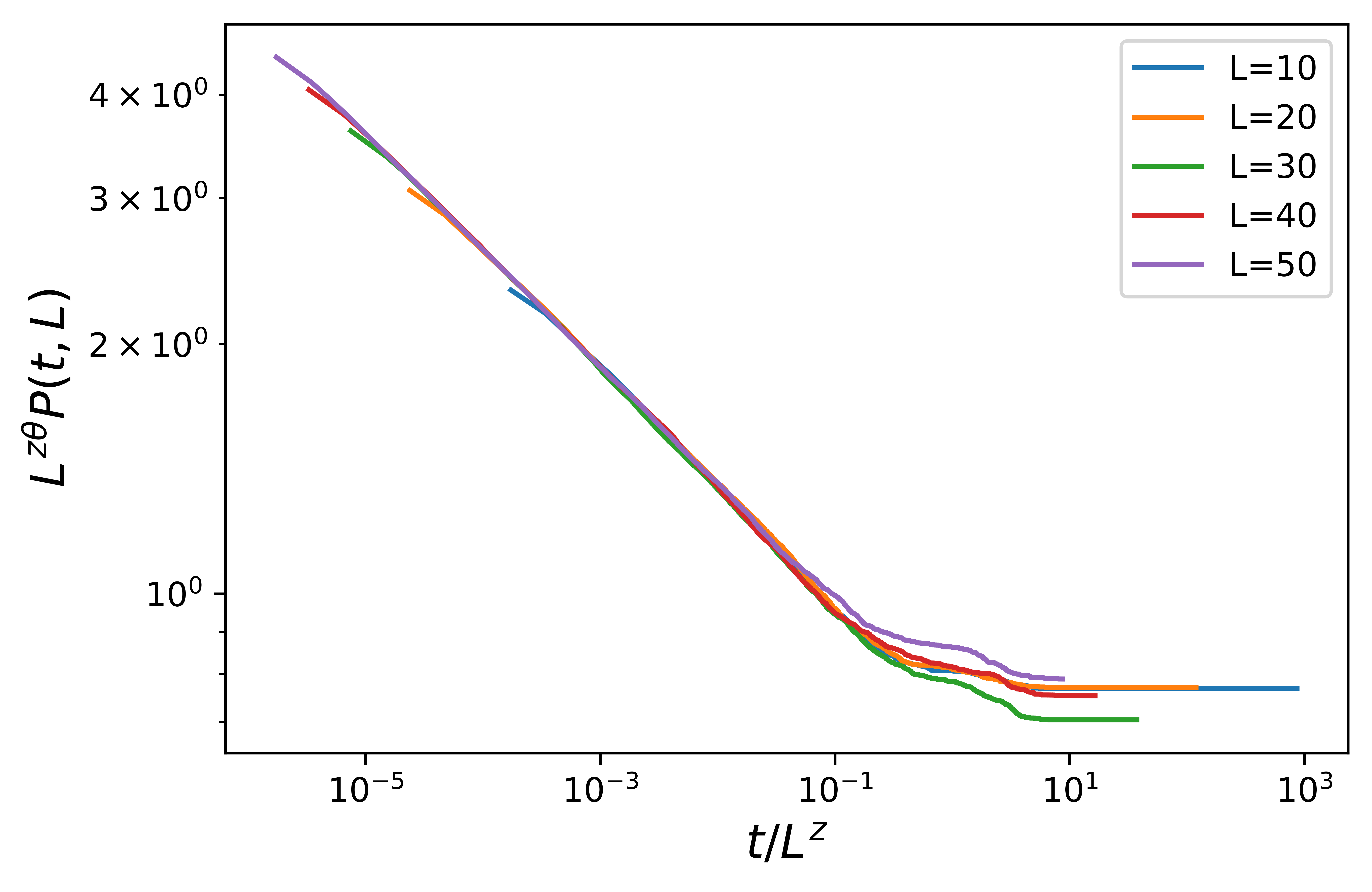}
		\caption{}
		\label{fig:2d_scaling_phi_0_5}
	\end{subfigure}
	~ 
	\caption{(a) Finite-size effects on 2D disordered lattices with $\phi=0.5$ (at the percolation threshold). (b) scaling behavior of 2D disordered lattices for $\phi = 0.5$ (at the percolation threshold) with $\theta = 0.141$ and $z = 2.86$.}
\end{figure*}

\begin{figure*}
	\centering
		\centering
	\begin{subfigure}[l]{0.45\linewidth}
		\includegraphics[width=\textwidth]{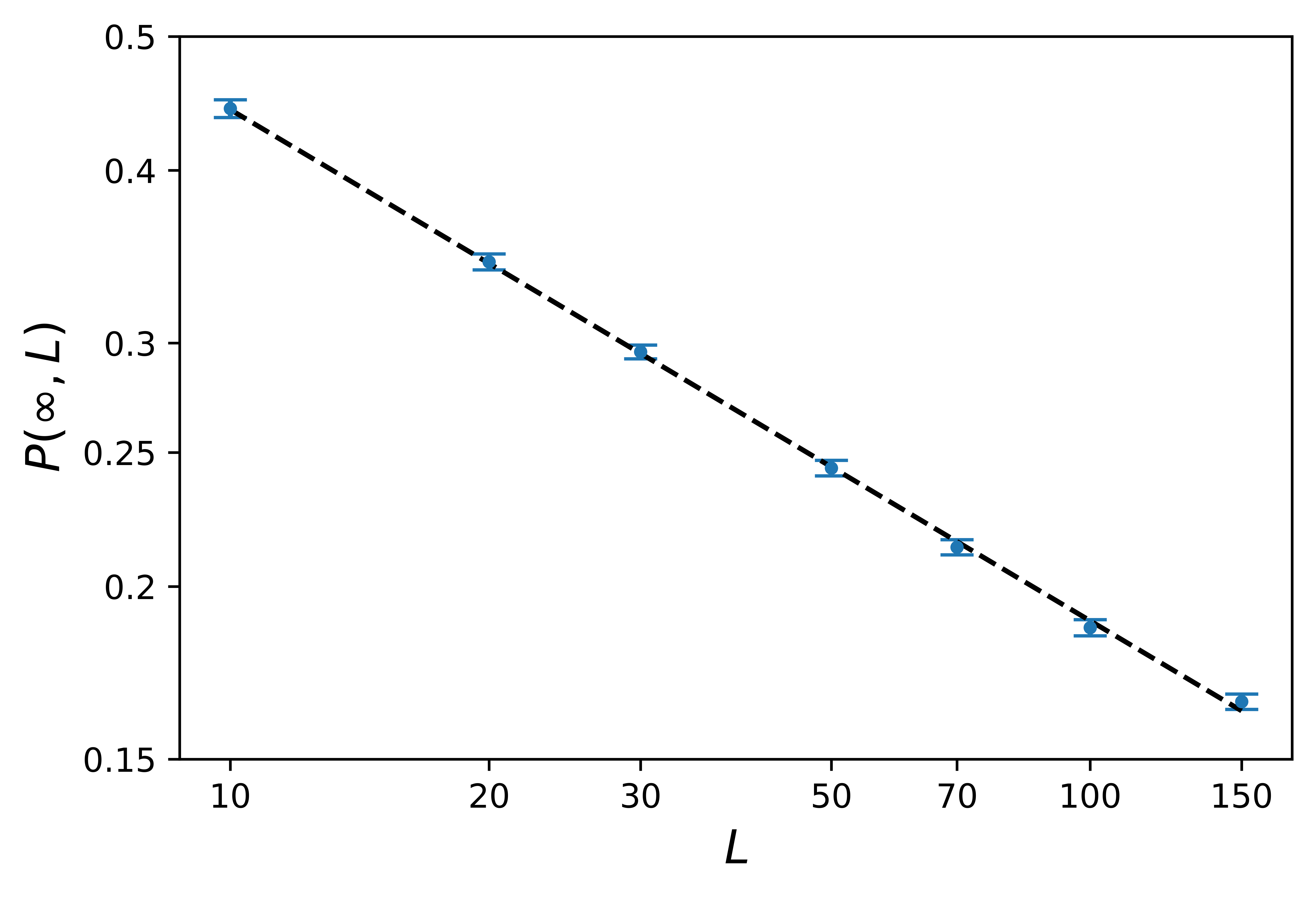}
		\caption{}
		\label{fig:2d_p_inf_1}
	\end{subfigure}
	\begin{subfigure}[l]{0.45\linewidth}
		\includegraphics[width=\textwidth]{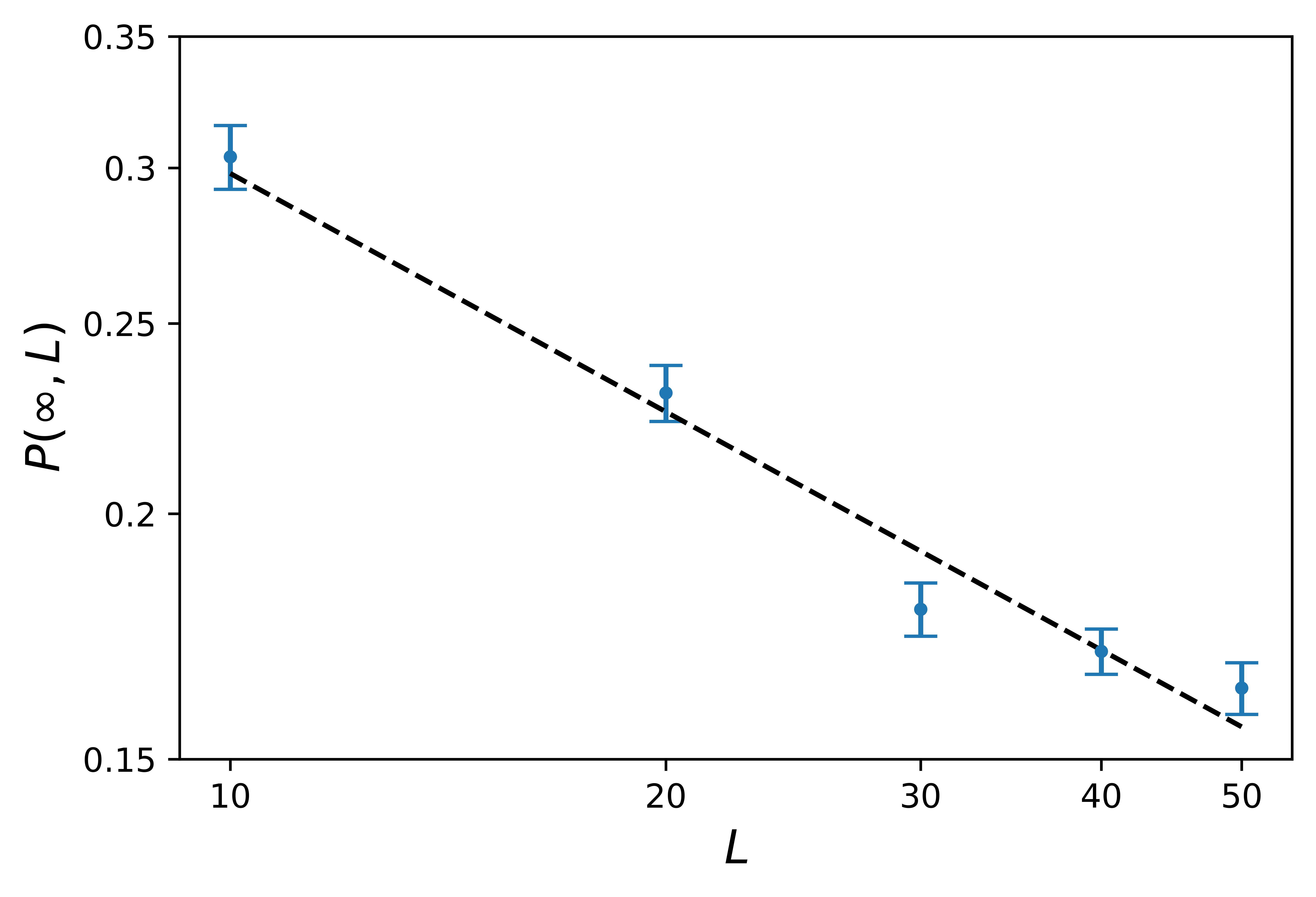}
		\caption{}
		\label{fig:2d_p_inf_0_5}
	\end{subfigure}
	\caption{The long-time asymptotic value, $P(\infty, L)$ of the 2D disordered lattice with (a) $\phi = 1$ and (b) $\phi = 0.5$ (at the percolation threshold) obeys a power-law relation as a function of the system length, $L$. The error bars show the standard error over 100 runs. The dotted black line is the best-fit power-law for $P(\infty, L) \propto L^{-z\theta}$.}\label{fig:lim_val_vs_L}
\end{figure*}

Fig.~\ref{fig:2d_fin_size_1} shows that there are characteristic finite-size effects in the persistence curves where they flatten at some non-zero value. We refer to this limiting value as $P(\infty, L)$, and the crossover time at which this limiting value is reached as $t_{\times}$. For simple diffusion on 2D regular lattices, this crossover times scales as $t_{\times}\sim L^z$, with $z=2$ \cite{havlin1983,manoj2000,Majumdar1996}.

On lattices of length $L$, the limiting value of persistence has a power-law behavior $P(\infty, L) \propto L^{-z\theta}$~\cite{manoj2000}, where $\theta$ is the persistence exponent associated with the temporal power-law decay of $P(t, L)$ and $z$ is a (dynamic) scaling exponent. Following previous work~\cite{havlin1983,manoj2000,Majumdar1996}, we take the scaling behavior of the persistence probability to be of the form
\begin{align}
    P(t, L) &= L^{-z\theta}f(t/L^z),
\end{align}
where
\begin{align}
    f(x) &=
\begin{cases}
      x^{-\theta} & \text{ if } x \ll 1, \\
      \text{constant} & \text{ if } x \gg 1.
\end{cases}
\end{align}
We obtain $z$ by plotting $P(\infty,L)$ vs. $L$ and fitting a power law with the exponent $-z\theta$, as shown in Fig.~\ref{fig:2d_p_inf_1}. For $\phi = 1$ we obtain $z = 1.99 \pm 2.8 \times 10^{-3}$, in good agreement with the literature value of the scaling exponent~\cite{manoj2000}. The scaling behavior of 2D lattices with $\phi = 1$ is shown in Fig.~\ref{fig:2d_scaling_phi_1} and we see good collapse of all the curves.

We then also observe similar finite-size effects for disordered lattices, as shown in Fig.~\ref{fig:2d_fin_size_0_9} for $\phi = 0.9$. Using the same value of $z\simeq 1.99$ that we determined for $\phi = 1$, we plot the scaling behavior in Fig.~\ref{fig:2d_scaling_phi_0_9}.
While with increasing system size the collapse of the scaled persistence curves onto a single curve improves, corrections to scaling are considerably stronger than for the regular 2D lattice.

When the 2D disordered lattice is at the percolation threshold $\phi_{c} = 0.5$ we observe that approaching the asymptotic values of the persistence curves appear at a much longer time scales as can be seen in Fig.~\ref{fig:2d_fin_size_phi_0_5}. Employing the same method that we used previously, we find $z = 2.86\pm 3.2 \times 10^{-2}$ for $\phi=0.5$ (Fig.~\ref{fig:2d_p_inf_0_5}). Fig.~\ref{fig:2d_scaling_phi_0_5} displays $L^{z\theta}P(t, L)$ vs. $t/L^z$. We can see from the figure that there is reasonably good collapse of the different curves onto a single scaling function. Note, however, that the combination of slow relaxation and large sample-to-sample fluctuations give rise to significantly larger error bars in the values of the scaling exponents.

\subsection*{Erd\H{o}s-R\'enyi Networks}

For Erd\H{o}s-R\'enyi networks \cite{ER_1960} studied in this and the following sections, for numerical stability, we use $\Delta t = \frac{1}{100}$ as we consider larger average degrees $\langle k \rangle $ than for the 2D bond-percolation networks above.

\begin{figure*}
	\centering
	\begin{subfigure}[l]{0.45\linewidth}
		\includegraphics[width=\textwidth]{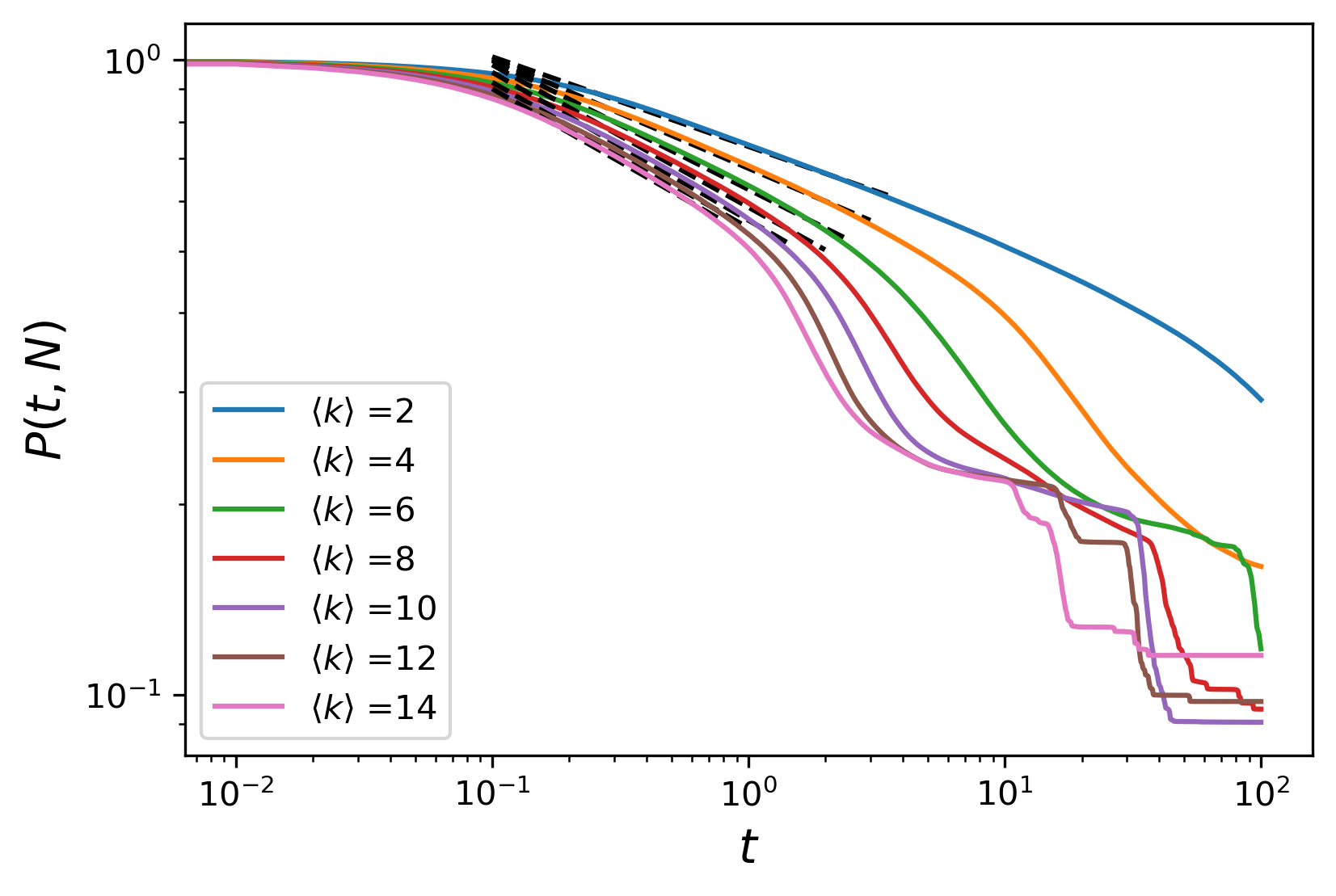}
		\caption{}
		\label{fig:er_per_k_comp}
	\end{subfigure}
	\begin{subfigure}[l]{0.45\linewidth}
		\includegraphics[width=\textwidth]{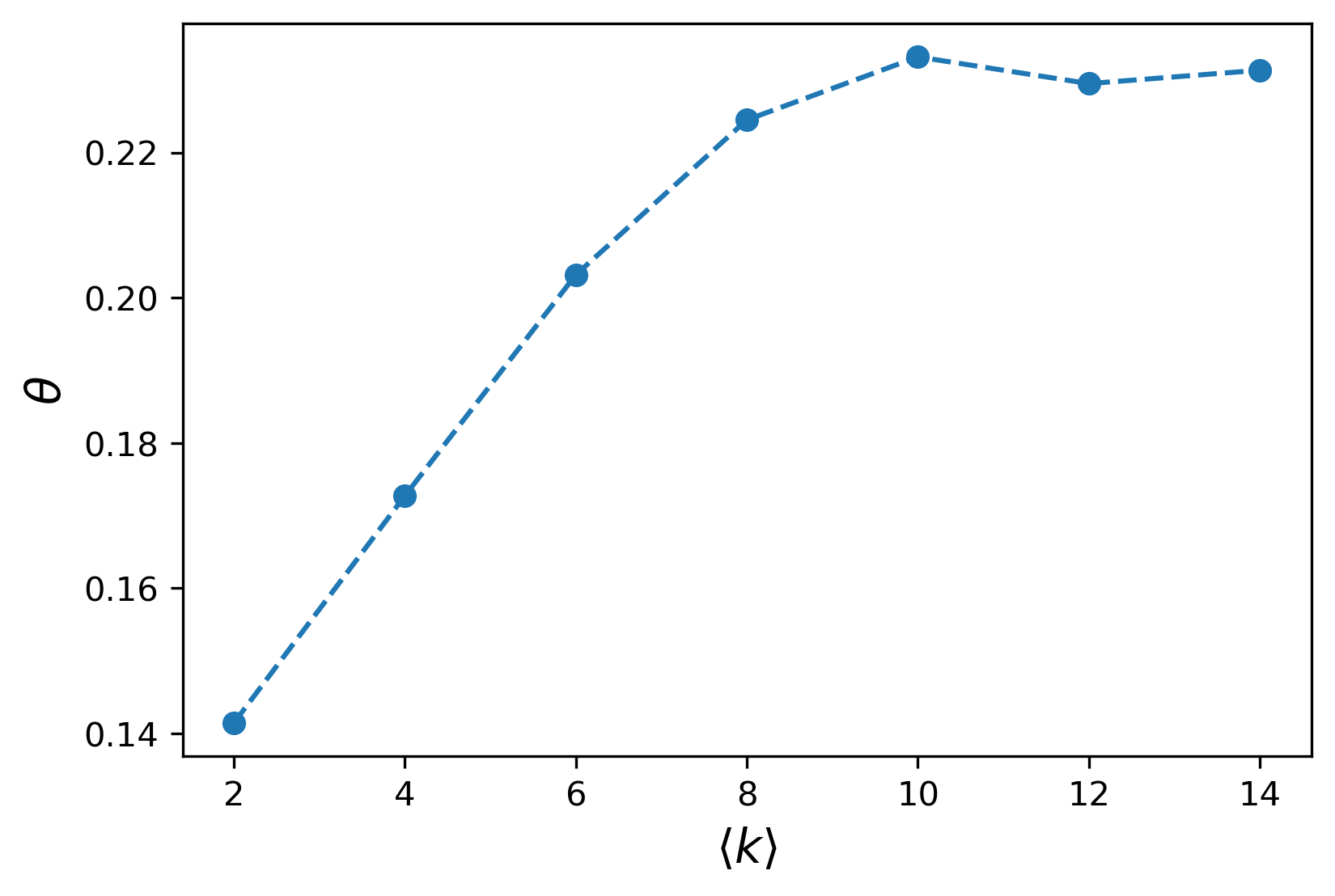}
		\caption{}
		\label{fig:er_theta_vs_k}
	\end{subfigure}
	~ 
	\caption{(a) Local persistence for ER networks with $N=10^4$. The dotted lines show the segment that has been fited with a power-law. (b) The power-law exponent of the persistence curves as a function of the average degree. }\label{fig:er_per}
\end{figure*}

\begin{figure*}
	\centering
	\begin{subfigure}[l]{0.45\linewidth}
		\includegraphics[width=\textwidth]{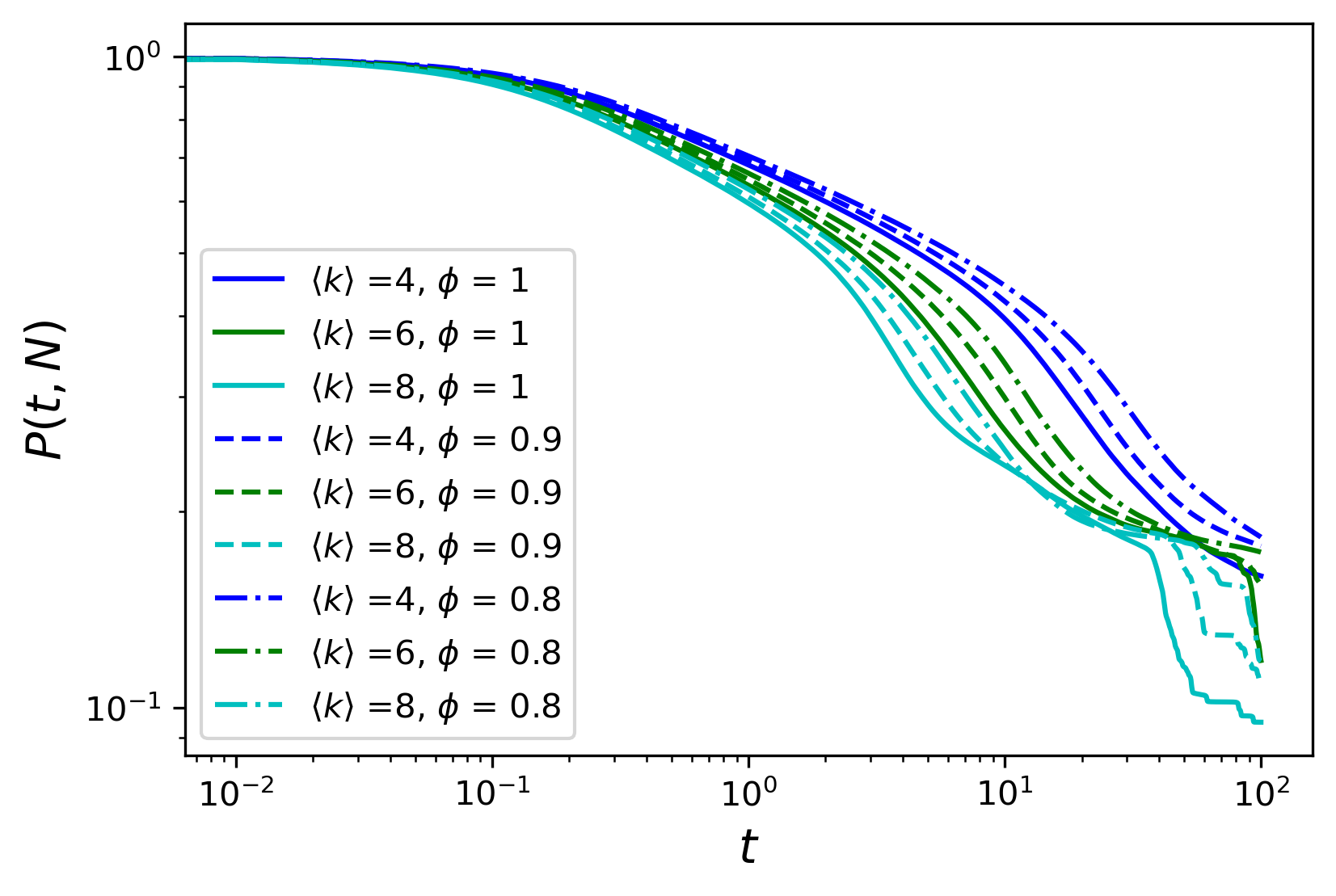}
		\caption{}
		\label{fig:er_phi_comp}
	\end{subfigure}
	\begin{subfigure}[r]{0.45\linewidth}
		\includegraphics[width=\textwidth]{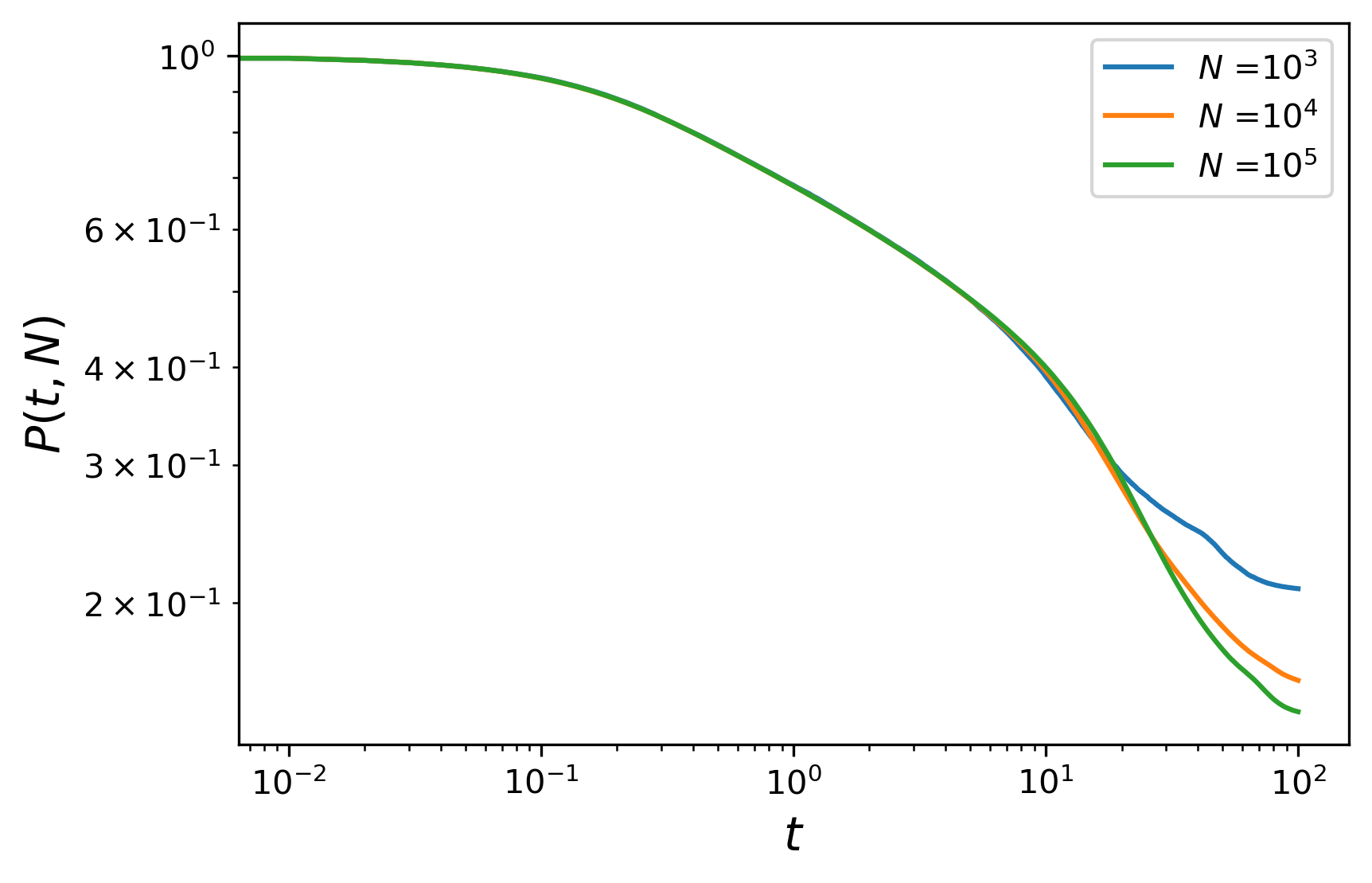}
		\caption{}
		\label{fig:er_size_comp}
	\end{subfigure}
	~ 
	\caption{(a) Diffusive persistence on Erd\H{o}s-R$\acute{\text{e}}$nyi networks ($N = 10^4$) for different average degrees, $\braket{k}$, and different edge removal probabilities, $1-\phi$. (b) Diffusive persistence for ER networks of different sizes with $\braket{k}=4$.}
\end{figure*}
\begin{figure*}
	\centering
	\begin{subfigure}[l]{0.45\linewidth}
		\includegraphics[width=\textwidth]{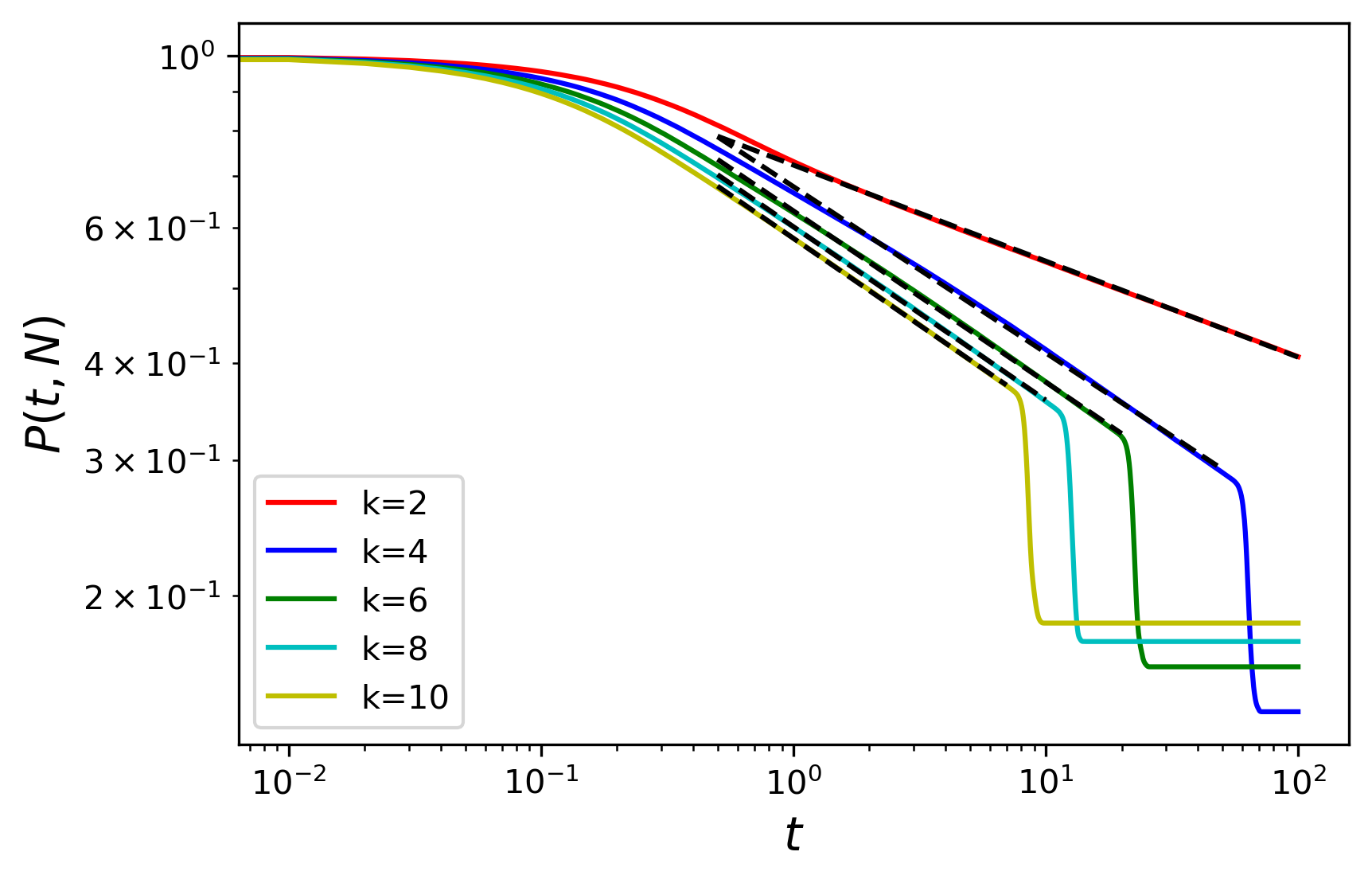}
		\caption{}
		\label{fig:reg_per_k_comp}
	\end{subfigure}
	\begin{subfigure}[l]{0.45\linewidth}
		\includegraphics[width=\textwidth]{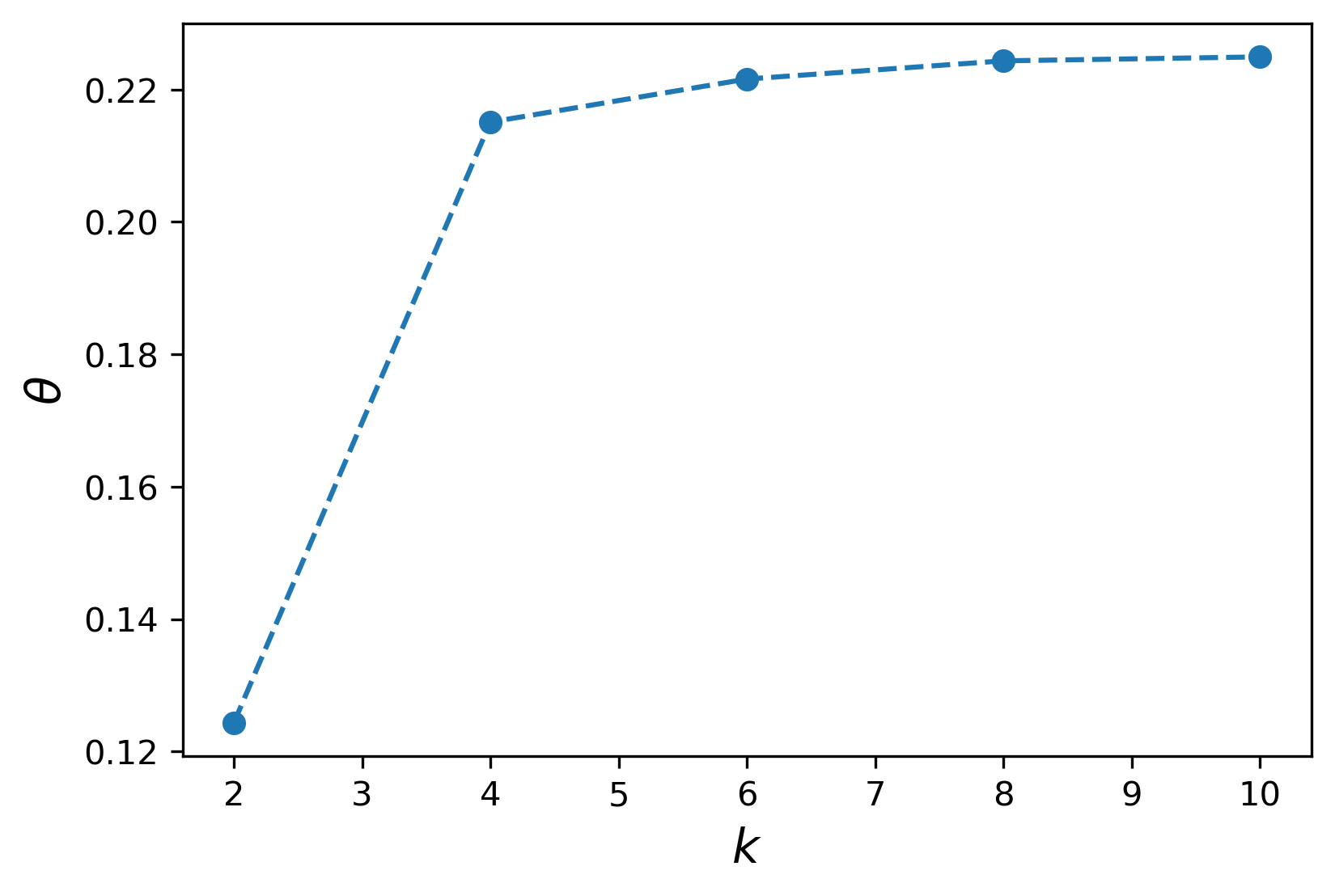}
		\caption{}
		\label{fig:reg_theta_vs_k}
	\end{subfigure}
	~ 
	\caption{(a) Local persistence for $k$-regular random networks with $N=10^4$. The dotted lines show the segment that has been fitted with a power-law. (b) The power-law exponent of the persistence curves as a function of the network degree. }\label{fig:reg_per}
\end{figure*}

\begin{figure*}
	\centering
	\begin{subfigure}[l]{0.45\linewidth}
		\includegraphics[width=\textwidth]{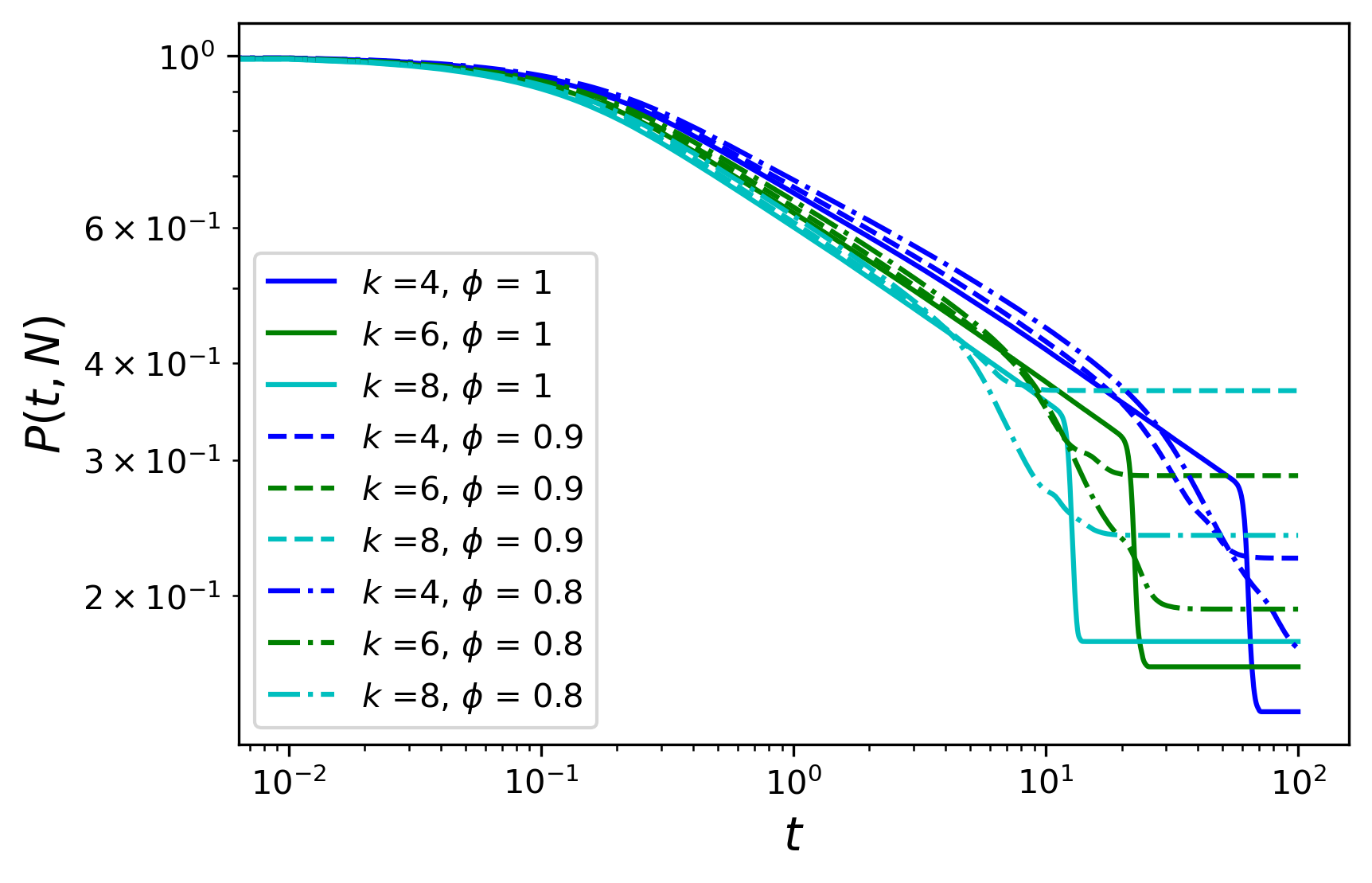}
		\caption{}
		\label{fig:reg_phi_comp}
	\end{subfigure}
	\begin{subfigure}[r]{0.45\linewidth}
		\includegraphics[width=\textwidth]{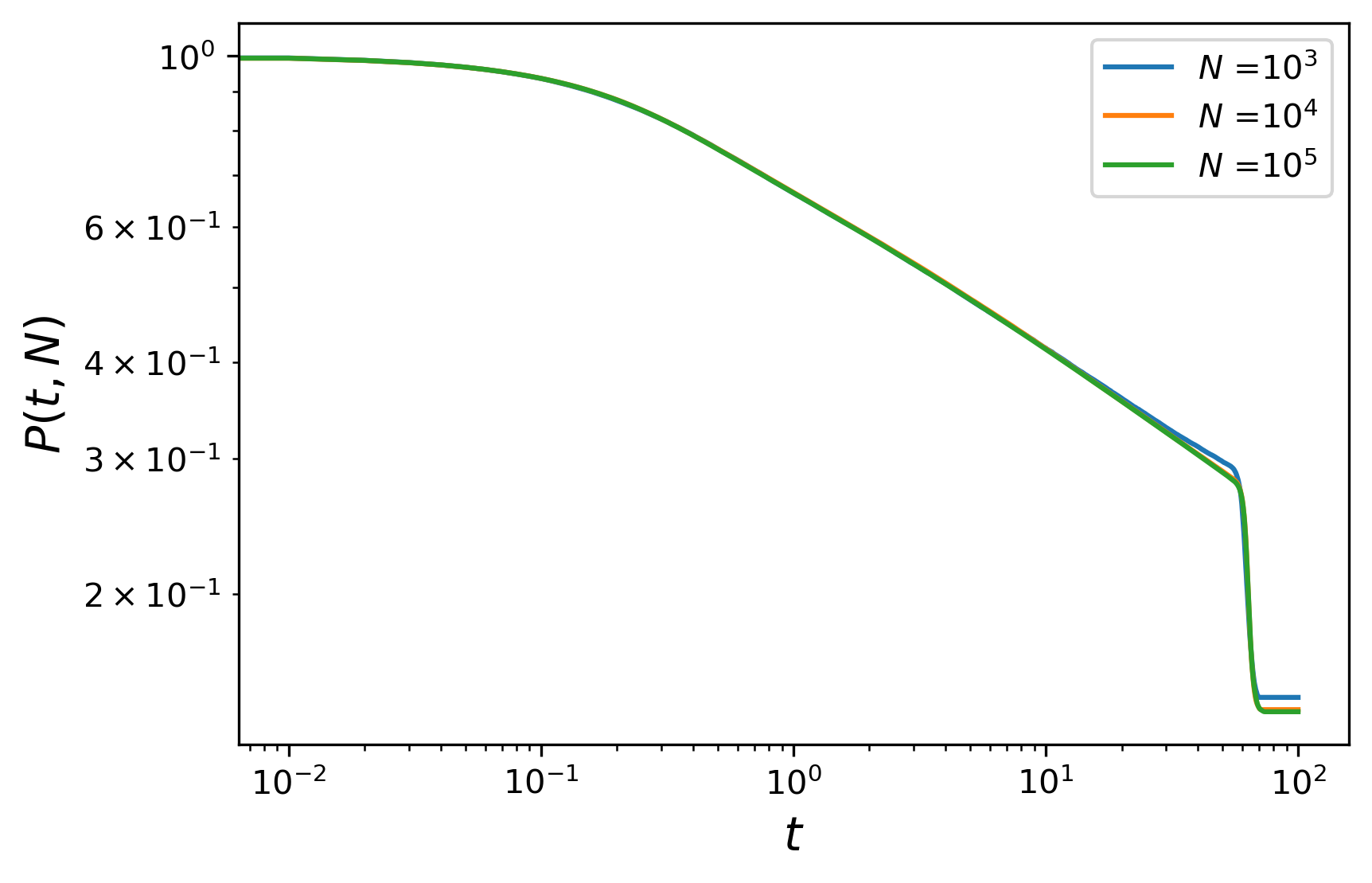}
		\caption{}
		\label{fig:reg_size_comp}
	\end{subfigure}
	~ 
	\caption{(a) Diffusive persistence on $k$-regular random networks ($N = 10^4$) for different $k$, and different edge removal probabilities, $1-\phi$. (b) Diffusive persistence for $k$-regular random networks of different sizes with $k=4$.}
\end{figure*}

The behavior of persistence is shown in Fig.~\ref{fig:er_per_k_comp}. In contrast to the clear power-law scaling of 2D networks, no such behavior is observed for ER networks. While we do not see any clear power-law scaling, we do notice that as the average degree of the network increases, the persistence curves cluster increasingly close to each other. By fitting a power-law to a segment of the persistence curves (marked by the dotted lines in Fig.~\ref{fig:er_per_k_comp}) we are able to characterize this effect, as shown in Fig.~\ref{fig:er_theta_vs_k}.

Given a realization of an ER network, we then proceed similarly as we did in the case of 2D lattices, by removing edges from the network with probability $1-\phi$. Here $\langle k \rangle $ denotes the average degree of the ER graphs before edge removal.
Fig.~\ref{fig:er_phi_comp} shows persistence curves for different average degrees and different values of $\phi$. As more edges are removed from a network of a given average degree we see that its persistence behavior changes to that of a network with a smaller average-degree. This is not surprising, since removing a certain fraction of edges from an ER network does not cause its topology to change, rather only its degree distribution is changed.

Fig.~\ref{fig:er_size_comp} shows that the particular shape of the curves is not defined by the network size but is a function of the particular network topology.

The persistence curves for $\braket{k}\ge 5$ begin to flatten out, showing that the value of the persistence drops steeply. ER networks exhibit fast and efficient relaxation, and the changes in the values of the field quickly become smaller than the numerical precision of the variables that store the field values.

\subsection*{$k$-Regular Random Networks}

The behavior of diffusive persistence for $k$-regular random networks is shown in Fig.~\ref{fig:reg_per_k_comp}. We see a distinct segment of the curve that shows power-law scaling. Similar to ER networks, we notice that as $k$ increases the persistence curves become closer, and by plotting the power-law exponent $\theta$ as a function of $k$ in Fig.~\ref{fig:reg_theta_vs_k}, we see that this exponent reaches an asymptotic value of approximately $0.22$ as $k$ becomes larger.

Fig.~\ref{fig:reg_phi_comp} shows that as we delete edges from the network, the persistence maintains its power-law behavior for some segment of the curve.

Similar to the ER case, we see that the values of the persistence curves steeply drop after a certain period of time. As before, this effect is due to the rapid relaxation of the field, causing changes in the values of the field to quickly become smaller than the numerical precision of the variables that hold those values.

\subsection*{$k$-th power of the ring}

\begin{figure}
	\centering
	\includegraphics[width=\linewidth]{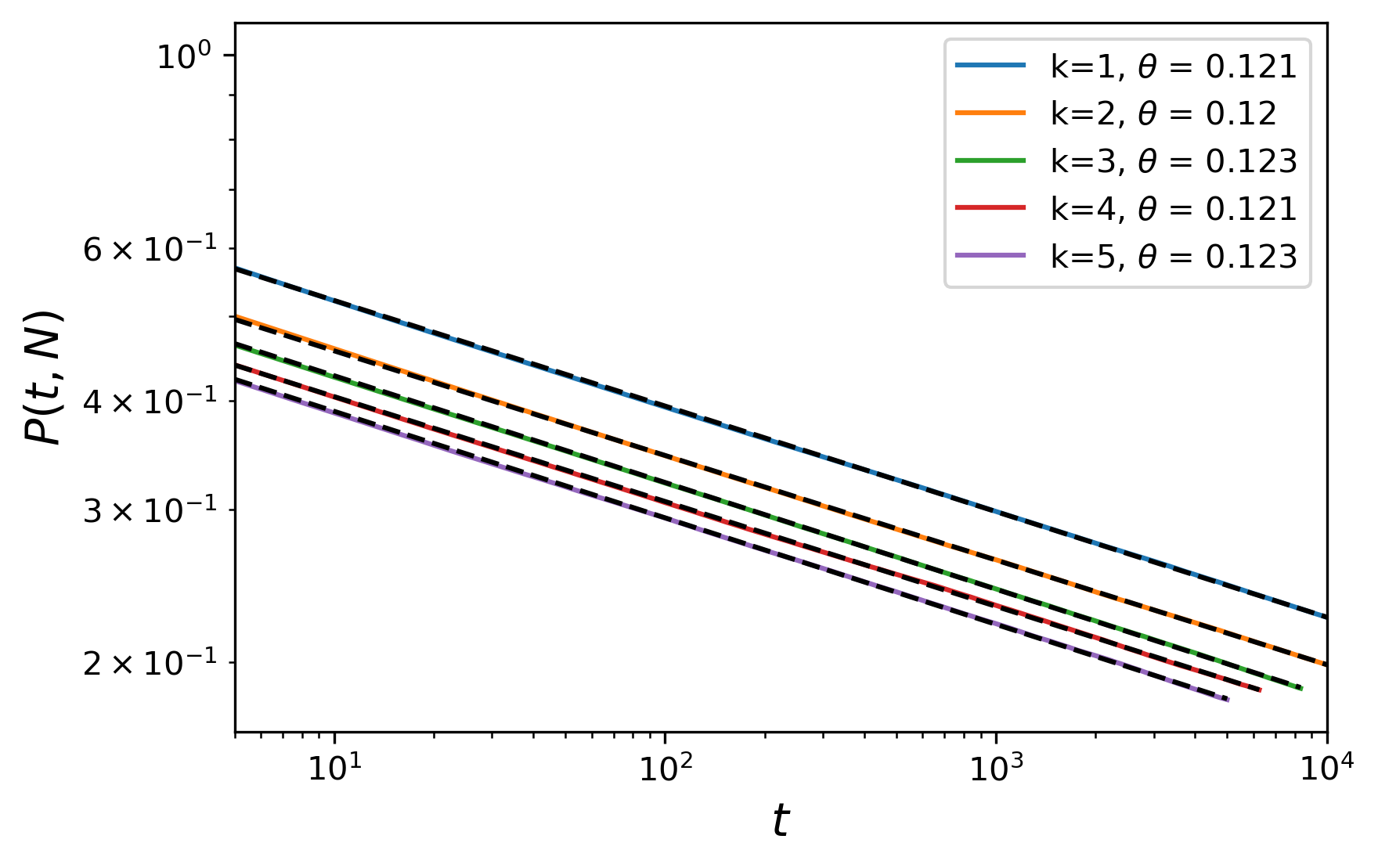}
	\caption{Diffusive persistence for the $k$-ring with 1000 nodes. The dotted lines show the power-law fit with the exponent $\theta$}\label{fig:k_ring}
\end{figure}

These graphs are regular, with a ring structure, such that every node has $k$ consecutive neighbors to the right and $k$ consecutive neighbors to the left and thus, it is also translationally invariant. Clearly, $k=1$ is simply just a cycle or 1D ring or lattice. For a network on $N$ nodes where $N = 2p+1$, $k=p$ will generate the complete graph $K_{N}$, where all pairs of nodes are connected with a link. The DPP for $k = 1$ behaves as the power-law \eqref{pow} with the exponent  $\theta_{1} = 0.1207...$, whereas for $k = (N-1)/2$ (the complete graph), probability $P(t)$ is a constant $0.5$ \cite{ZTNotes}. This implies that as $k$ increases from a constant to its maximum value, the persistence behavior also changes.
Figure~\ref{fig:k_ring} shows the decay of $P(t,N)$ versus $t$ for increasing values of $k$.


\subsection*{Random geometric graphs in 2D}

\begin{figure*}
	\centering
	\begin{subfigure}[l]{0.45\linewidth}
		\includegraphics[width=\textwidth]{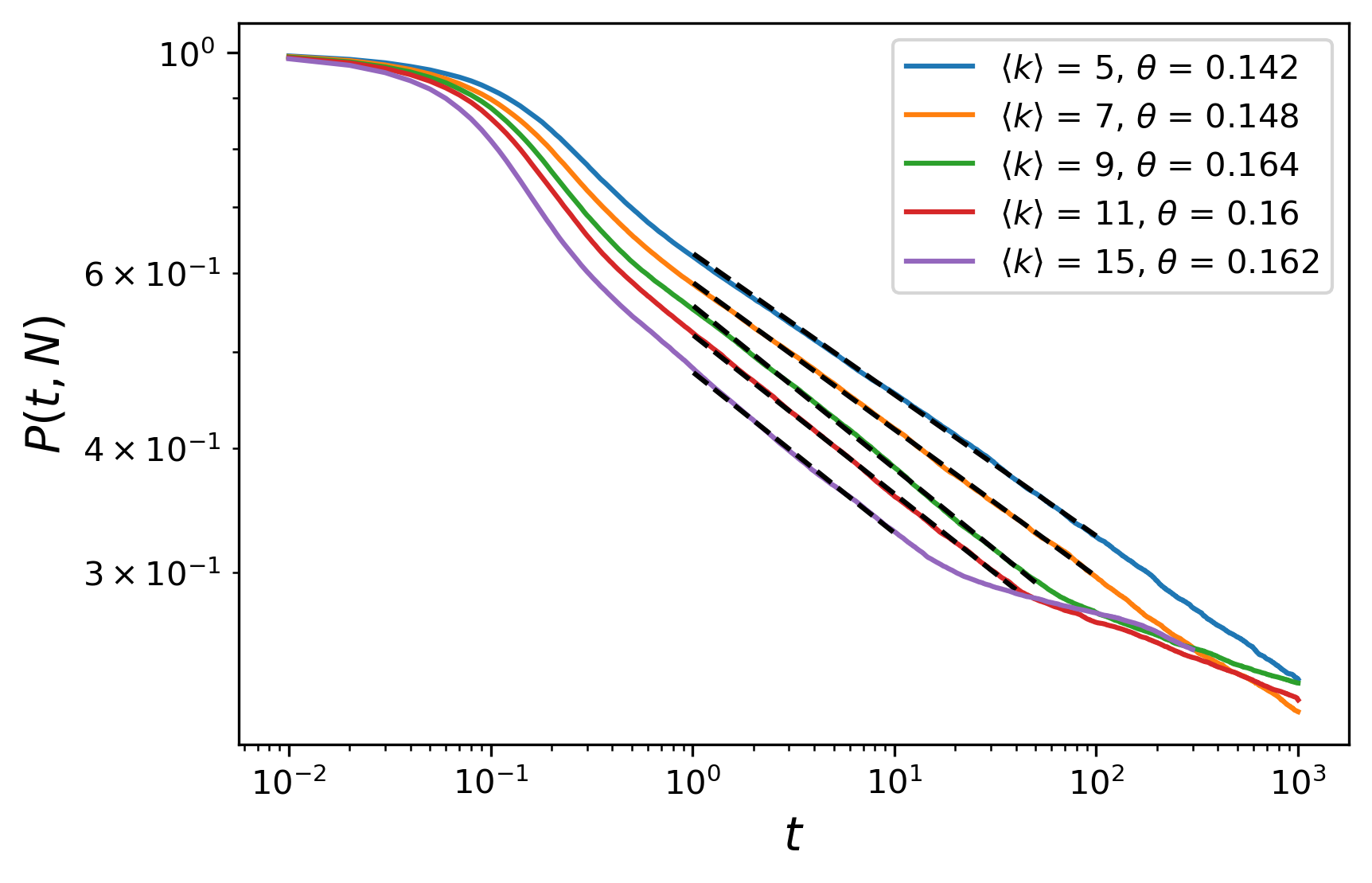}
		\caption{}
		\label{fig:rgg_2d_low}
	\end{subfigure}
	\begin{subfigure}[r]{0.45\linewidth}
		\includegraphics[width=\textwidth]{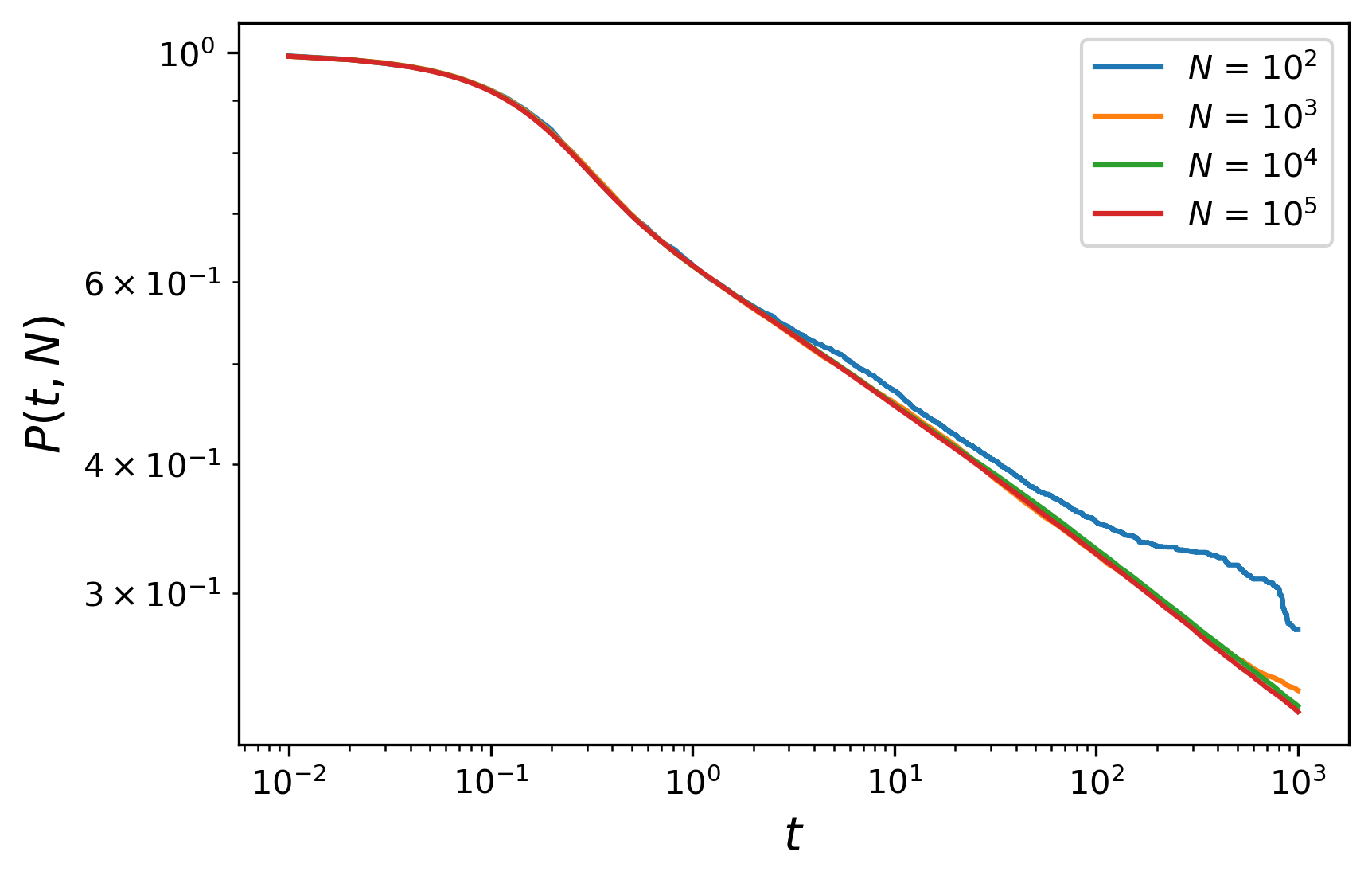}
		\caption{}
		\label{fig:rgg_2d_high}
	\end{subfigure}
	\caption{(a) Diffusive persistence on 2D random geometric graphs with $N = 10^3$. The dotted lines indicate the power-law fit with the exponent $\theta$. As the average degree of the networks increases, the value of the power-law exponent saturates. (b) Diffusive persistence on 2D random geometric graphs of different sizes with average degree $\braket{k} = 5$. No finite size-effects appear for graphs with more than $10^2$ nodes in the timescale we studied.}
	\label{fig:rgg_2d}
\end{figure*}

Random geometric graphs (RGG) \cite{Penrose_RGG,Dall_PRE2002} in 2D are obtained by sprinkling $N$ points \balance uniformly at random into the unit square $[0,1]^{2}$, then connecting every pair of points that are within a prescribed (Euclidean) distance $R$. The random network created that way will have an average degree of $\langle k \rangle = \pi R^{2} N$ and a Poisson degree distribution just like the ER graphs. However, unlike the ER graphs, RGG-s are spatially embedded, have a high clustering coefficient and have no shortcuts. RGG-s are similar to graphs generated by continuous percolation.
Increasing $R$, the average degree increases and at a critical value $\langle k \rangle_{c} = 4.52\pm 0.01$ (in 2D) a giant connected cluster appears that spans the unit square. Figure~\ref{fig:rgg_2d} shows the evolution of $P(t,N)$ versus $t$ for different $\langle k \rangle$ values on the largest connected component of the RGG. It also shows power-law scaling behavior with exponents that are smaller than the $\theta = 0.1875$ value for the 2D lattice,
but with slowly increasing exponents. Again, if $R \geq \sqrt{2}$, the RGG becomes a complete graph with a constant
value for the DPP, which implies that there has to be a drastic change in the behavior of the persistence probability as the graphs are getting denser.

\section*{Discussions}

Persistence problems are rather meaningful and relevant in complex networks not just from a theoretical, but also a practical view, as described in the Introduction section. However, due to their typically non-uniform nature, translational invariance no longer holds and we have to distinguish between local and global persistence. The former studies persistence at the node level, whereas the latter at the whole network level, which was the focus of this paper. We have shown that the properties of diffusive persistence  are highly non-trivial even at the global level and in general, they depend on network parameters. Our results are purely numerical in nature but we hope that they are interesting enough to initiate analytical studies in the future.

Our investigations show that in regular networks, such as 2D square lattices, the persistence probability obeys clear power-law scaling. As the network becomes more disordered through the inverse bond percolation process, the power-law exponent remains constant at $\theta \simeq 0.186$ for $\phi > \phi_c$ but with some strong corrections to the scaling behavior. However, at $\phi_c$ we observe the novel power-law exponent $\theta \simeq 0.141$. This change may likely be associated with the phase-transition that the network topology goes through, at the percolation threshold.
We also studied finite-size effects at the percolation threshold and discovered that the DPP $P(t,L)$ and the finite-size crossover times exhibit dynamical scaling with exponent $z \simeq 2.86$, different from the known value of $z = 2$ on 2D regular lattices.

We also observe clear power-law behavior for $k$-regular random networks, however, we observe no such clear scaling behavior for ER networks.

\section*{Acknowledgements}
  O.M., A.M., D.H., B.S., and G.K. were supported in part by the Defense Advanced Research Projects Agency (DARPA) and the Army Research Office (ARO) under Contract No. W911NF-17-C-0099, and the Army Research Laboratory (ARL) under Cooperative Agreement Number W911NF-09-2-0053 (NS-CTA). Z.T. was supported in part by the NSF Grant No. IIS-1724297, by the Defense Threat Reduction Agency Award No. HDTRA1-09-1-0039, and jointly by the U.S. Air Force Office of Scientific Research (AFOSR) and the Defense Advanced Research Projects Agency (DARPA) under contract FA9550-12-1-0405. The views and conclusions contained in this document are those of the authors and should not be interpreted as representing the official policies either expressed or implied of the Army Research Laboratory or the U.S. Government.

\newpage


\newpage

\bibliographystyle{plain}
\bibliography{persistence-bib}

\end{document}